\documentclass[conference]{IEEEtran}

\IEEEoverridecommandlockouts

\usepackage{hyperref}
\usepackage{graphicx}
\usepackage[caption=false,font=footnotesize]{subfig}
\usepackage{bibentry}
\usepackage{cite}
\usepackage{amsmath,amssymb,amsfonts,amssymb}
\usepackage{algorithmic}

\graphicspath{{./graphics/}}
\usepackage{textcomp}
\usepackage{xcolor}
\def\BibTeX{{\rm B\kern-.05em{\sc i\kern-.025em b}\kern-.08em
    T\kern-.1667em\lower.7ex\hbox{E}\kern-.125emX}}
\usepackage{siunitx}
\usepackage{tabularx}
\usepackage{url}
\usepackage{booktabs}
\usepackage{diagbox}
\usepackage{multirow}
\usepackage{soul}

\usepackage{cuted}

\usepackage{pgfplots}
\pgfplotsset{compat=newest}
\usepgfplotslibrary{fillbetween}
\usepgfplotslibrary{external}
\tikzexternaldisable
\newboolean{snr}

\definecolor{color0}{rgb}{0.12156862745098,0.466666666666667,0.705882352941177}
\definecolor{color1}{rgb}{1,0.498039215686275,0.0549019607843137}
\definecolor{color2}{rgb}{0.172549019607843,0.627450980392157,0.172549019607843}
\definecolor{color3}{rgb}{0.83921568627451,0.152941176470588,0.156862745098039}
\definecolor{color4}{rgb}{0.580392156862745,0.403921568627451,0.741176470588235}
\definecolor{gray}{rgb}{0.75,0.75,0.75}

\newlength\figureheight
\newlength\figurewidth
\newlength\confheight
\newlength\confwidth
\newlength\axisheight
\newlength\axiswidth
\usepackage[colorinlistoftodos,prependcaption,textsize=tiny]{todonotes}

\newif\ifpreprint
\preprinttrue

\begin{document}


\title{The Importance of Being Earnest: Performance of Modulation Classification for Real RF Signals\\
}

\author{\IEEEauthorblockN{Colin de Vrieze, Ljiljana Simi\'c, Petri M\"ah\"onen}
\IEEEauthorblockA{Institute for Networked Systems \\
RWTH Aachen University \\
Email: colin.de.vrieze@rwth-aachen.de, \{lsi, pma\}@inets.rwth-aachen.de}
}

\maketitle


\begin{abstract}
Digital modulation classification (DMC) can be highly valuable for equipping radios with increased spectrum awareness in complex emerging wireless networks. However, as the existing literature is overwhelmingly based on theoretical or simulation results, it is unclear how well DMC performs in practice. In this paper we study the performance of DMC in real-world wireless networks, using an extensive RF signal dataset of 250,000 over-the-air transmissions with heterogeneous transceiver hardware and co-channel interference. Our results show that DMC can achieve a high classification accuracy even under the challenging real-world conditions of modulated co-channel interference and low-grade hardware. However, this only holds if the training dataset fully captures the variety of interference and hardware types in the real radio environment; otherwise, the DMC performance deteriorates significantly. Our work has two important engineering implications. First, it shows that it is not straightforward to exchange learned classifier models among dissimilar radio environments and devices in practice. Second, our analysis suggests that the key missing link for real-world deployment of DMC is designing signal features that generalize well to diverse wireless network scenarios. We are making our RF signal dataset publicly available as a step towards a unified framework for realistic DMC evaluation.

\end{abstract}


\section{Introduction}
Digital modulation classification (DMC) is a subtype of signal classification concerned with the recognition of the underlying digital modulation scheme from received radio signal samples. DMC is becoming highly relevant not only for military, but also civilian applications. 
As wireless networks become more complex, DMC can be a valuable tool for increased spectrum awareness~\cite{kim2007} and more efficient spectrum access algorithms~\cite{tsakmalis2014}, by affording the radio a more sophisticated understanding of its environment. For example, in unlicensed spectrum it is advantageous for radios to know which technology they are sharing the spectrum with~\cite{hu2014}, whereas in dynamic spectrum access scenarios, DMC can help secondary users differentiate between primary and secondary transmissions based on their signal characteristics~\cite{clancy2011}.

In their seminal work ~\cite{azzouz1995,azzouz1996,nandi1998}, Azzouz and Nandi introduced different DMC algorithms based on now widely-adopted spectral features extracted from the received signals. Subsequent works introduced new, high-order statistics features~\cite{soliman1992,swami2000,orlic2012}. There have since been a large number of publications exploring variants of DMC with different sets of features and algorithms; for a detailed survey see~\cite{hazza2013,dobre2007}. However, the vast majority of the literature is based on either theoretical or simulation results. It thus remains unclear how DMC performs in real-world wireless networks, given the influence of transceiver hardware on signal integrity and the channel impairments of a real radio environment.

The authors in~\cite{swami2000} present a thorough theoretical analysis of cumulant features, but their evaluation is based on simulations with coherent and synchronous nodes, which greatly simplifies the practical conditions of a wireless communication link. Real-world challenges like multipath and interference are addressed in~\cite{orlic2012}, but their evaluation is likewise purely simulation-based and does not consider the widely adopted spectral features from~\cite{azzouz1996}. The authors in \cite{nandi2004} consider a complex artificial neural network classifier but also assume a simplified channel model and synchronization among nodes. Simpler support vector machines are used for extensive DMC simulations in~\cite{gang2004} and~\cite{llenas2017}. To the best of our knowledge,~\cite{llenas2017} is one of the rare works to include a limited over-the-air hardware evaluation of their DMC classifier.

The existing studies of DMC performance thus do not properly capture the richness of the operational environment for DMC in real wireless networks, being based solely on simulation results or a narrow hardware demonstration. Importantly, practical constraints such as the influence of transceiver hardware or co-channel interference have remained largely unaddressed in the literature. Our work provides three main contributions towards closing the gap between simulation-based performance evaluations of DMC in the literature and the reality of wireless networks. First, we provide classification accuracy results on a comprehensive and diversified RF signal dataset of \num[group-separator={,}]{250000} real-world over-the-air transmissions, rather than using simulated signals. Second, we study the effect of modulated co-channel interference on DMC performance, as a key component of a real radio environment for DMC applications. Finally, we consider a diverse set of transceiver hardware to capture the effects that arise from the heterogeneity of real radio systems, and which have implications for the feasibility of exchanging learned classifier models among devices in practice. 

The rest of the paper is organized as follows. Section~\ref{sec:classification} presents the considered classification algorithm and the underlying features. Section~\ref{sec:setup} presents our scenarios, measurement setup, and evaluation methodology. Section~\ref{sec:results} presents our results. Section~\ref{sec:conclusion} concludes the paper.

\section{SVM-based Signal Classification}\label{sec:classification}

Given a two-class classification problem with a set of sample feature vectors $(x_i)$ and labels $y_i\in\{-1,1\}$ indicating their individual class membership, a linear support vector machine (SVM)~\cite{abe2010} seeks to find a hyperplane $w^Tx_i+b$ in the feature space that best separates the samples according to their class. Any new sample can then be classified based on the location of its feature vector on either side of the separating hyperplane. In case of data that is not linearly separable, a soft-margin SVM introduces slack variables to tolerate occasional transgressing samples, and the kernel trick can enable nonlinear decision boundaries. The binary SVM can be extended to differentiate between more than two classes by combining the results of multiple one-vs-rest classifiers.

In this work, we use a multiclass soft-margin SVM to classify signals of different modulation based on spectral and statistical features extracted from raw recorded I/Q samples. We intentionally selected a simple linear SVM over more complex alternatives like ANN as our aim is to study the fundamental performance of DMC on real-world RF signals; we expect that a different classifier would not change our overall conclusions. We used the SVC class from the ``sklearn'' Python library with slack parameter $c=10$. Table~\ref{tab:features} lists the considered features, also widely adopted in literature~\cite{azzouz1995, hazza2013}.

\begin{table}[tb!]
\caption{Features extracted from the recorded RF signals}
\begin{center}
\begingroup
	\setlength{\tabcolsep}{2pt} 
	\begin{tabularx}{\columnwidth}{l X}
	\toprule
	\textbf{Feature}     & \textbf{Description} (definition as in cited reference or in Appendix) \\
	\midrule
	$\hat{SNR}$	 	 	 & Estimated signal-to-noise ratio (\ref{eq:snr})\\
	$\sigma_{aa}$        & Std. dev. of abs. centered and normalized (CN) amplitude \cite{azzouz1996}\\
	$\sigma_{a}$         & Std. dev. of CN amplitude over non-weak (NW) I/Q samples \cite{azzouz1996}\\
	$\sigma_{af}$        & Std. dev. of abs. CN instant. frequency over NW I/Q samples \cite{azzouz1996}\\
	$N_c$                & Fraction of NW I/Q samples in signal (\ref{eq:nw})\\
	$\tilde{C}_{42}$ & Normalized, noise corrected fourth-order cumulant (\ref{eq:c42})\\
	$\tilde{C}_{63}$ & Normalized, noise corrected sixth-order cumulant (\ref{eq:c63})\\
	$\gamma_{2,m}$       & Max. value of DFT of squared CN amplitude \cite{hazza2013}\\
	$\gamma_{4,m}$       & Max. value of DFT of 4th power CN amplitude \cite{hazza2013} \\
	$\mu_{42}^{f}$       & Kurtosis of CN instant. frequency over NW I/Q samples \cite{azzouz1996}\\
	$R_{xx,m}$           & Maximum of autocorrelation function \\
	$PAPR$        		 & Peak-to-average power ratio \\
	\hline
	\end{tabularx}
\endgroup
\label{tab:features}
\end{center}
\end{table}

\section{Experiment Scenarios \& Methodology}\label{sec:setup}


\subsection{Scenarios}\label{sec:scenarios}

\subsubsection{Baseline Scenario}\label{scen:baseline}
We evaluate the baseline performance of the classifier over a range of signal-to-noise (SNR) values. To this end, we generate high-integrity training and testing datasets using laboratory-grade hardware, i.e. an Agilent 81180A arbitrary waveform generator with an Agilent E4438C vector signal generator as transmitter, and an Agilent N9030A spectrum analyzer as receiver. 

\subsubsection{Low-Cost Hardware Scenario}\label{scen:sdr}

We study the effect of low-cost transceiver hardware on DMC performance. To this end, we use USRP B200 software defined radios (SDRs) as both transmitter and receiver. This scenario allows us to investigate the impact of low signal integrity, as SDRs have lower sensitivity, higher nonlinearities, and greater frequency and DC offsets than our baseline laboratory-grade devices. We also mix training and testing datasets generated by the two radio types to study how well the signal features generalize with respect to the heterogeneous hardware. The results from this experiment have practical implications for whether learned models can be exchanged between dissimilar device types.

\subsubsection{Co-Channel Interference Scenario}\label{scen:cci}

We study the effect of interference on DMC performance. Modulated co-channel interference is arguably the most challenging for DMC since the statistics of the interfering modulated signal might confuse the classifier.  Similarly, we consider the most challenging hardware configuration, classifying a low-integrity signal (i.e. SDR as transmitter) against a high-integrity interfering signal (i.e. laboratory-grade transmitter as interferer) using a low-integrity SDR receiver. 
We consider a range of signal-to-interference ratios (SIR) and different interferer modulations. We also study DMC performance when trained solely on non-interfered signals but tested in the presence of interference.

%
%

\subsection{RF Signal Data Generation}
Our measurements were taken in a small indoor office environment using dipole antennas at \SI{5.75}{GHz}.  We consider the modulation types of BPSK, QPSK, 16-QAM and 64-QAM for single carrier (SC) modulation and orthogonal frequency division multiplex (OFDM). For SC signals, we generate symbols from random data at a rate of \SI{10}{Msps}, then resample the signal at \SI{40}{MHz} and use a root-raised-cosine pulse-shaping filter. For OFDM, we adopt the IEEE~802.11a PHY symbol design using 64 subcarriers, $11$ guards, empty zero carrier, a cyclic prefix with relative length of $0.25$, and the same set of subcarrier modulation schemes as for SC; OFDM signals with different modulations are subsequently aggregated into a \emph{single} OFDM class.  We randomize the pilots between $\{-1, 1\}$ and oversample the IFFT by a factor of $2$ to approximately achieve the same spectral bandwidth of \SI{20}{MHz} as the SC signals. To make the classification task more challenging, we use no preambles throughout.

Before recording the signals, we first coarsely correct the carrier frequency offset between transmitter and receiver down to a maximum deviation of \SI{100}{Hz}, and, if applicable, randomize the carrier frequency offset of the interferer within $\pm\SI{5}{ppm}$. Then, the SNR and SIR at the receiver are calibrated to the desired values by tuning the power of the transmitting radios using spectral density estimates from the receiver's sampled data. 
Once the transmitter and, if applicable, the interferer have been calibrated, the receiver samples \SI{1}{ms} long signal segments at \SI{50}{MHz} I/Q sample rate to achieve an oversampling factor of $2.5$. The receiver extracts an estimate of the SNR, as defined in (\ref{eq:snr}) in the Appendix, before low-pass filtering the signal to suppress any noise outside of the \SI{20}{MHz} signal bandwidth for feature extraction. We do not equalize or correct any residual frequency offset, sampling clock offsets, or inter-symbol interference from the pulse-shaping filter, so that our results generalize to a blind receiver. Finally, we featurize each signal segment by extracting the features listed in Table~\ref{tab:features} and save those values in our database.

\subsection{Performance Evaluation Methodology} 

We select all the featurized recordings from our signal database (comprising \num[group-separator={,}]{250000} segments in total) that apply to the given scenario (\textit{cf.} Section~\ref{sec:scenarios}). Using this scenario dataset, we run a ten-fold shuffle split cross-validation of DMC performance as follows. First, we randomly sample disjunct training and testing datasets, each comprising \num[group-separator={,}]{10000} featurized segments. Before training the classifier, we standardize both datasets according to the mean and standard deviation of the training set. We train the SVM~(\textit{cf.} Section~\ref{sec:classification}) using the training dataset and save the decision boundaries. We then evaluate the classification accuracy, defined as the percentage of signal segments in the test dataset correctly identified according to their modulation scheme. This is repeated ten times; the cross-validation allows us to avoid sampling bias and report a variance estimate alongside the mean classification accuracy.

\section{Results \& Analysis}\label{sec:experiments} \label{sec:results}

\subsection{Baseline Performance}

\begin{table}[tb!]
	\centering
	\caption{Classification accuracy for the baseline scenario, based on training with data at all SNR levels}
	\label{tab:baseline}
	\begingroup
		\setlength{\tabcolsep}{2pt} 
		\begin{tabularx}{\columnwidth}{c>{\centering\arraybackslash}X>{\centering\arraybackslash}X>{\centering\arraybackslash}X>{\centering\arraybackslash}X>{\centering\arraybackslash}X c}
		\toprule
		\textbf{SNR}                    & \SI{0}{dB}        & \SI{5}{dB}        & \SI{10}{dB}       & \SI{15}{dB}       & \SI{20}{dB}       & Avg. accuracy \\
		\midrule
		\textbf{Accuracy}               & 90.8\%            & 98.8\%            & 100.0\%           & 100.0\%           & 100.0\%           & 97.9\%   \\[-0.5ex]
		\tiny($\pm$ std. dev.) & \tiny$(\pm0.8\%)$ & \tiny$(\pm0.2\%)$ & \tiny$(\pm0.0\%)$ & \tiny$(\pm0.0\%)$ & \tiny$(\pm0.0\%)$ & \tiny$(\pm0.2\%)$ \\
		\bottomrule
		\end{tabularx}
	\endgroup
\end{table}

\tikzsetnextfilename{conf_baseline}%
\begin{figure}[tb!]
	\centering
		\def\datafile{tikz/01_baseline.dat}
\vspace{-0.5cm}
\begin{tikzpicture}
\begin{axis}[
    width={\confwidth},
    height=\confheight,
    colorbar,
    colormap={mymap}{
        rgb(0pt)=(0.83921568627451,0.152941176470588,0.156862745098039);
        rgb(1pt)=(1,1,1);
        rgb(2pt)=(0.172549019607843,0.627450980392157,0.172549019607843)
    },
    colorbar style={
        ytick={-1,0,1},
        yticklabels={\raisebox{2\height}{\parbox{1cm}{100\% incorrect}}, not \mbox{assigned}, \raisebox{-2.5\height}{\parbox{1cm}{100\% correct}}},
        ylabel={},
        tick pos=right,
        tick align=outside,
        text width=1cm,
        font=\footnotesize,
    },
    xlabel={Predicted class},
    ylabel={True class},
    xlabel style={
        yshift=5,
    },
    xmin=-0.5, xmax=4.5,
    ymin=-0.5, ymax=4.5,
    xtick={0,1,2,3,4},
    ytick={0,1,2,3,4},
    xticklabels={BPSK,QPSK,16QAM,64QAM,OFDM},
    yticklabels={BPSK,QPSK,16QAM,64QAM,OFDM},
    xticklabel style = {
        rotate=45,
        font=\scriptsize,
        anchor=east,
        xshift=1,
        yshift=-1,
    },
    yticklabel style = {
        rotate=45,
        font=\scriptsize,
        xshift=1,
        yshift=2,
        anchor=east,
    },
    xlabel style = {font=\footnotesize},
    ylabel style = {font=\footnotesize},
    tick pos=left,
    tick align=outside,
    axis equal image,
    point meta min=-1,
    point meta max=1,
    font=\small,
]

    \addplot [
        matrix plot,
        mesh/cols=5,
        point meta=\thisrow{score},
        nodes near coords={
            \scriptsize \pgfmathprintnumber[fixed, fixed zerofill, precision=2]{\numscore} 
            \tiny (\pgfmathprintnumber[set thousands separator={}]{\numcorr})
        },
        nodes near coords align={anchor=center},
        nodes near coords style={
            text width=0.75cm,
            align=center,
            font=\tiny,
            },
        visualization depends on={\thisrow{num_corr} \as \numcorr},
        visualization depends on={abs(\thisrow{score}) \as \numscore},
    ] table {\datafile};
\end{axis}
\end{tikzpicture}
	\caption{Confusion matrix for the \textit{baseline} scenario in Table~\ref{tab:baseline} over all SNR levels, for a single cross-validation fold (absolute numbers in brackets).}
	\label{fig:baseline}
\end{figure}
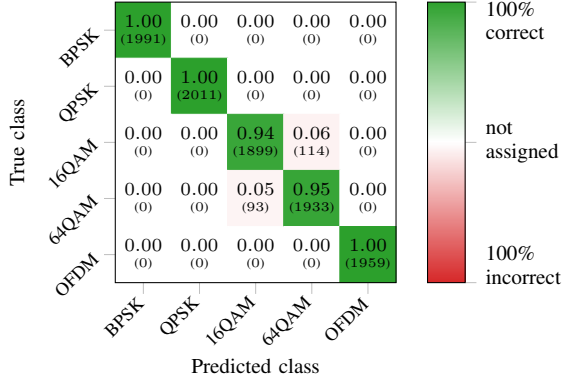
\tikzexternaldisable

\tikzexternalenable
\begin{figure}[t]
	\centering
	\setlength\figureheight{0.4\columnwidth}%
	\setlength\figurewidth{0.60\columnwidth}%
	\subfloat{

\begin{tikzpicture} 
    \begin{axis}[%
        hide axis,
        xmin=0, xmax=1,
        ymin=0, ymax=1,
        font=\footnotesize,
        legend columns=5,
        legend style={draw=none, column sep=1pt},
    ]
    \addlegendimage{color0, thick}
    \addlegendimage{color1, thick}
    \addlegendimage{color2, thick}
    \addlegendimage{color3, thick}
    \addlegendimage{color4, thick}
    \addlegendentry{BPSK}
    \addlegendentry{QPSK}
    \addlegendentry{16-QAM}
    \addlegendentry{64-QAM}
    \addlegendentry{OFDM}
    \end{axis}
\end{tikzpicture}%
	}%
	\vspace{-0.2cm}\setcounter{subfigure}{0}
	%
	\def\xlabel{Standardized feature value}
	\def\datafile{tikz/kde/5db/kde_nc.dat}
	\tikzsetnextfilename{kde_nc_5}%
	\subfloat[][Distribution of $N_c$ values]{%
		\begin{minipage}[t]{0.48\columnwidth}%
			\centering%

\begin{tikzpicture} 

\begin{axis}[
    set layers,
    every axis plot/.append style={on layer=pre main},
    width=\figurewidth,
    height=\figureheight,
    xlabel={\xlabel},
    xlabel style={yshift=3},
    ylabel={Probablity density},
    ymin=0, 
    ymajorticks=false,
    xticklabel style={font=\scriptsize},
    clip mode=individual,
    font=\footnotesize,
    enlarge x limits=0.02,
]


\addplot[
        color=color0,
        draw opacity=0.75,
        fill=color0,
        fill opacity=0.5,
        thick,
] table [x=xs,y=BPSK]{\datafile}\closedcycle;
\addplot[
        color=color1,
        draw opacity=0.75,
        fill=color1,
        fill opacity=0.5,
        thick,
] table [x=xs,y=QPSK]{\datafile};
\addplot[
        color=color2,
        draw opacity=0.75,
        fill=color2,
        fill opacity=0.5,
        thick,
] table [x=xs,y=16QAM]{\datafile};

\addplot[
        color=color3,
        draw opacity=0.75,
        fill=color3,
        fill opacity=0.5,
        thick,
] table [x=xs,y=64QAM]{\datafile};
\addplot[
        color=color4,
        draw opacity=0.75,
        fill=color4,
        fill opacity=0.5,
        thick,
] table [x=xs,y=OFDM]{\datafile};
\end{axis} 
\end{tikzpicture}%
		\end{minipage}%
	}\hfill%
	\def\datafile{tikz/kde/5db/kde_sigmaa.dat}%
	\tikzsetnextfilename{kde_sigmaa_5}%
	\subfloat[][Distribution of $\sigma_{a}$  values]{
		\begin{minipage}[t]{0.48\columnwidth}
			\centering

\begin{tikzpicture} 

\begin{axis}[
    set layers,
    every axis plot/.append style={on layer=pre main},
    width=\figurewidth,
    height=\figureheight,
    xlabel={\xlabel},
    xlabel style={yshift=3},
    ylabel={Probablity density},
    ymin=0, 
    ymajorticks=false,
    xticklabel style={font=\scriptsize},
    clip mode=individual,
    font=\footnotesize,
    enlarge x limits=0.02,
]


\addplot[
        color=color0,
        draw opacity=0.75,
        fill=color0,
        fill opacity=0.5,
        thick,
] table [x=xs,y=BPSK]{\datafile}\closedcycle;
\addplot[
        color=color1,
        draw opacity=0.75,
        fill=color1,
        fill opacity=0.5,
        thick,
] table [x=xs,y=QPSK]{\datafile};
\addplot[
        color=color2,
        draw opacity=0.75,
        fill=color2,
        fill opacity=0.5,
        thick,
] table [x=xs,y=16QAM]{\datafile};

\addplot[
        color=color3,
        draw opacity=0.75,
        fill=color3,
        fill opacity=0.5,
        thick,
] table [x=xs,y=64QAM]{\datafile};
\addplot[
        color=color4,
        draw opacity=0.75,
        fill=color4,
        fill opacity=0.5,
        thick,
] table [x=xs,y=OFDM]{\datafile};
\end{axis} 
\end{tikzpicture}%
		\end{minipage}
	}

	\def\datafile{tikz/kde/5db/kde_gamma2.dat}
	\tikzsetnextfilename{kde_gamma2_5}%
	\subfloat[][Distribution of $\gamma_{2,m}$  values]{%
		\begin{minipage}[t]{0.48\columnwidth}%
			\centering%

\begin{tikzpicture} 

\begin{axis}[
    set layers,
    every axis plot/.append style={on layer=pre main},
    width=\figurewidth,
    height=\figureheight,
    xlabel={\xlabel},
    xlabel style={yshift=3},
    ylabel={Probablity density},
    ymin=0, 
    ymajorticks=false,
    xticklabel style={font=\scriptsize},
    clip mode=individual,
    font=\footnotesize,
    enlarge x limits=0.02,
]


\addplot[
        color=color0,
        draw opacity=0.75,
        fill=color0,
        fill opacity=0.5,
        thick,
] table [x=xs,y=BPSK]{\datafile}\closedcycle;
\addplot[
        color=color1,
        draw opacity=0.75,
        fill=color1,
        fill opacity=0.5,
        thick,
] table [x=xs,y=QPSK]{\datafile};
\addplot[
        color=color2,
        draw opacity=0.75,
        fill=color2,
        fill opacity=0.5,
        thick,
] table [x=xs,y=16QAM]{\datafile};

\addplot[
        color=color3,
        draw opacity=0.75,
        fill=color3,
        fill opacity=0.5,
        thick,
] table [x=xs,y=64QAM]{\datafile};
\addplot[
        color=color4,
        draw opacity=0.75,
        fill=color4,
        fill opacity=0.5,
        thick,
] table [x=xs,y=OFDM]{\datafile};
\end{axis} 
\end{tikzpicture}%
		\end{minipage}%
	}\hfill%
	\def\datafile{tikz/kde/5db/kde_c42.dat}%
	\tikzsetnextfilename{kde_c42_5}%
	\subfloat[][Distribution of $\tilde{C}_{42}$ values]{
		\begin{minipage}[t]{0.48\columnwidth}
			\centering

\begin{tikzpicture} 

\begin{axis}[
    set layers,
    every axis plot/.append style={on layer=pre main},
    width=\figurewidth,
    height=\figureheight,
    xlabel={\xlabel},
    xlabel style={yshift=3},
    ylabel={Probablity density},
    ymin=0, 
    ymajorticks=false,
    xticklabel style={font=\scriptsize},
    clip mode=individual,
    font=\footnotesize,
    enlarge x limits=0.02,
]


\addplot[
        color=color0,
        draw opacity=0.75,
        fill=color0,
        fill opacity=0.5,
        thick,
] table [x=xs,y=BPSK]{\datafile}\closedcycle;
\addplot[
        color=color1,
        draw opacity=0.75,
        fill=color1,
        fill opacity=0.5,
        thick,
] table [x=xs,y=QPSK]{\datafile};
\addplot[
        color=color2,
        draw opacity=0.75,
        fill=color2,
        fill opacity=0.5,
        thick,
] table [x=xs,y=16QAM]{\datafile};

\addplot[
        color=color3,
        draw opacity=0.75,
        fill=color3,
        fill opacity=0.5,
        thick,
] table [x=xs,y=64QAM]{\datafile};
\addplot[
        color=color4,
        draw opacity=0.75,
        fill=color4,
        fill opacity=0.5,
        thick,
] table [x=xs,y=OFDM]{\datafile};
\end{axis} 
\end{tikzpicture}%
		\end{minipage}
	}%
	\caption{Estimated probability densities of selected features for signal segments recorded at SNR=\SI{5}{dB},‚ for the \textit{baseline} scenario.}
	\label{fig:features}
\end{figure}
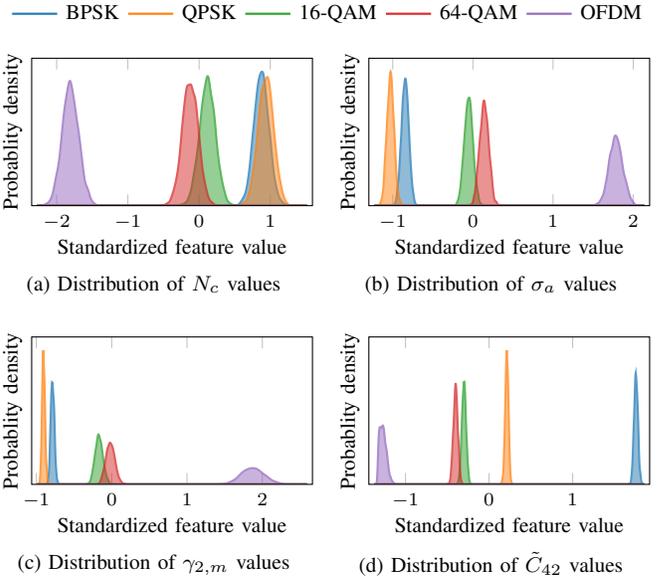
\tikzexternaldisable

Table~\ref{tab:baseline} shows the classifier performance versus SNR for the baseline scenario in Section~\ref{scen:baseline}, when the classifier has been trained on the full SNR range of $\{\SI{0}{dB}, \SI{5}{dB},\ldots,\SI{20}{dB}\}$. The results demonstrate a high baseline classification accuracy $(>\!90\%)$ with low variance for all SNR classes, and an average accuracy of $97.9\%$. To reveal the source of the misclassifications at low SNRs, Fig.~\ref{fig:baseline} shows a confusion matrix from a single representative cross-validation fold for the full SNR range. Fig.~\ref{fig:baseline} demonstrates that while all other modulations are classified perfectly, some confusion arises between 16-QAM and 64-QAM. This is also evident in Fig.~\ref{fig:features}, which shows the empirical distributions of selected features taken from a kernel density estimation (KDE)~\cite{parzen1962} of the feature values with a Gaussian kernel at $\text{SNR}=\SI{5}{dB}$. Fig.~\ref{fig:features} shows that the distributions for 16-QAM and 64-QAM partially overlap for all the features, which is consistent with the corresponding misclassifications\footnote{The underlying reason for misclassification is feature-dependent and a detailed discussion is out of the scope of this paper. However, the feature similarity of QAM modulation types is known from literature~\cite{swami2000,orlic2012}.} in Fig.~\ref{fig:baseline}. The same results, omitted for brevity, hold for $\text{SNR}=\SI{0}{dB}$.

\begin{table}[tb!]
	\centering
	\caption{Classification accuracy for the baseline scenario, based on training only with data at SNR=\SI{10}{dB}}
	\label{tab:snrtraining}
	\begingroup
		\setlength{\tabcolsep}{2pt} 
		\begin{tabularx}{\columnwidth}{c>{\centering\arraybackslash}X>{\centering\arraybackslash}X>{\centering\arraybackslash}X>{\centering\arraybackslash}X>{\centering\arraybackslash}X c}
		\toprule
		\textbf{SNR}           & \SI{0}{dB}         & \SI{5}{dB}         & \SI{10}{dB}       & \SI{15}{dB}       & \SI{20}{dB}        & Avg. accuracy \\
		\midrule
		\textbf{Accuracy}      & 39.8\%             & 60.5\%             & 100.0\%           & 58.0\%            & 57.2\%             & 63.1\%    \\[-0.5ex]
		\tiny($\pm$ std. dev.) & \tiny$(\pm12.7\%)$ & \tiny$(\pm10.9\%)$ & \tiny$(\pm0.0\%)$ & \tiny$(\pm5.8\%)$ & \tiny$(\pm10.6\%)$ & \tiny$(\pm6.8\%)$ \\
		\bottomrule
		\end{tabularx}
	\endgroup
\end{table}

Let us now consider how the classifier performs if only signals of a \textit{single} SNR class are available at training time.  
Table~\ref{tab:snrtraining} shows that if the classifier is only trained on data at $\text{SNR}=\SI{10}{dB}$, the classification accuracy across the other SNR classes degrades significantly, compared to when training over all SNR levels as in Table~\ref{tab:baseline}.  We note that similar results are obtained when training exclusively on data from any of the other SNR classes. It is tempting to attribute this performance degradation to the simplistic linear SVM and call for a more complex classifier such as ANN. However, we have established in Table~\ref{tab:baseline} that the classifier is capable of dealing with different noise levels, as long as the training data is representative of the test scenario. It is thus instructive to instead consider the behavior of the underlying features. Fig.~\ref{fig:dependency} highlights the difference in SNR dependency of two example features: $\tilde{C}_{42}$ shows no dependency on SNR, owing to the correction for noise power its formulation (\textit{cf.} (\ref{eq:c42})), whereas the value of $\sigma_{a}$ largely depends on the SNR during recording. Aside from $\tilde{C}_{42}$ and $\tilde{C}_{63}$, we have confirmed that all other features in Table~\ref{tab:features} are likewise SNR-dependent. Therefore, if the classifier only experiences signal segments of one SNR class during training, it is difficult to correctly assign the decision boundaries without a prior model of SNR dependency for each feature. We emphasize that the performance degradation thus appears to be due to feature design and \emph{not} due to a lack of inherent generalization of the classifier.

\tikzexternalenable
\begin{figure}[t!]
	\centering
		\setboolean{snr}{true}%
		\setlength\figureheight{0.5\columnwidth}%
		\setlength\figurewidth{0.53\columnwidth}%
		\subfloat{

\begin{tikzpicture} 
    \begin{axis}[%
        hide axis,
        xmin=0, xmax=1,
        ymin=0, ymax=1,
        font=\footnotesize,
        legend columns=5,
        legend style={draw=none, column sep=1pt},
    ]
    \addlegendimage{color0, thick, mark=x}
    \addlegendimage{color1, thick, mark=square}
    \addlegendimage{color2, thick, mark=o}
    \addlegendimage{color3, thick, mark=triangle}
    \addlegendimage{color4, thick, mark=diamond}
    \addlegendentry{BPSK}
    \addlegendentry{QPSK}
    \addlegendentry{16-QAM}
    \addlegendentry{64-QAM}
    \addlegendentry{OFDM}
    \end{axis}
\end{tikzpicture}%
		}%
		\vspace{-0.4cm}\setcounter{subfigure}{0}
		\def\ylabel{Standardized feature value}%
		\def\path{tikz/dependency/snr/sigma_a/}%
		\def\feature{$\sigma_{a}$}%
		\tikzsetnextfilename{dependency_snr_sigmaa}%
		\subfloat[][Dependency of $\sigma_{a}$ on SNR]{%
			\begin{minipage}[t]{0.5\columnwidth}
				\centering

\global\edef\bpsk{\path BPSK.csv}
\global\edef\qpsk{\path QPSK.csv}
\global\edef\qamst{\path 16QAM.csv}
\global\edef\qamsf{\path 64QAM.csv}
\global\edef\ofdm{\path OFDM.csv}

\ifthenelse{\boolean{snr}}
{
    \global\edef\labels{{20dB, 15dB, 10dB, 5dB, 0dB}}
    \global\edef\xlabel{SNR class}
    \global\edef\xmax{4}
}{
    \global\edef\labels{{20dB, 15dB, 10dB, 5dB}}
    \global\edef\xlabel{SIR class}
    \global\edef\xmax{3}
}

\begin{tikzpicture} 

\begin{axis}[
    set layers,
    every axis plot/.append style={on layer=pre main},
    width=\figurewidth,
    height=\figureheight,
    title style={yshift=-7},
    xlabel=\xlabel,
    ylabel=\ylabel,
    xtick = {0,1,2,3,4},
    xticklabels/.expanded = {\labels},
    xticklabel style={font=\scriptsize},
    yticklabel style={font=\scriptsize},
    ylabel style={yshift=-3pt},
    ytick={-5,-4,-3,-2,-1,0,1,2,3,4,5},
    xmin=0, xmax=\xmax,
    clip mode=individual,
    clip marker paths=true,
    font=\footnotesize,
]

\addplot[name path = A, draw=none, forget plot] 
    table [col sep=comma, x expr=\coordindex, y expr={\thisrow{mean}-1*\thisrow{std}}]{\bpsk};
\addplot[name path = B, draw=none, forget plot] 
    table [col sep=comma, x expr=\coordindex, y expr={\thisrow{mean}+1*\thisrow{std}}]{\bpsk};
\addplot [
    color=color0,
    opacity=0.25,
    forget plot,
    ] fill between [
    of=A and B
];
\addplot[
        color=color0,
        semithick,
        mark=x
] table [x expr=\coordindex,y=mean, col sep=comma]{\bpsk};

\addplot[name path = A, draw=none, forget plot] 
    table [col sep=comma, x expr=\coordindex, y expr={\thisrow{mean}-1*\thisrow{std}}]{\qpsk};
\addplot[name path = B, draw=none, forget plot] 
    table [col sep=comma, x expr=\coordindex, y expr={\thisrow{mean}+1*\thisrow{std}}]{\qpsk};
\addplot [
    color=color1,
    opacity=0.25,
    forget plot,
    ] fill between [
    of=A and B
];
\addplot[
        color=color1,
        semithick,
        mark=square
] table [x expr=\coordindex,y=mean, col sep=comma]{\qpsk};

\addplot[name path = A, draw=none, forget plot] 
    table [col sep=comma, x expr=\coordindex, y expr={\thisrow{mean}-1*\thisrow{std}}]{\qamst};
\addplot[name path = B, draw=none, forget plot] 
    table [col sep=comma, x expr=\coordindex, y expr={\thisrow{mean}+1*\thisrow{std}}]{\qamst};
\addplot [
    color=color2,
    opacity=0.25,
    forget plot,
    ] fill between [
    of=A and B
];
\addplot[
        color=color2,
        semithick,
        mark=o
] table [x expr=\coordindex,y=mean, col sep=comma]{\qamst};

\addplot[name path = A, draw=none, forget plot] 
    table [col sep=comma, x expr=\coordindex, y expr={\thisrow{mean}-1*\thisrow{std}}]{\qamsf};
\addplot[name path = B, draw=none, forget plot] 
    table [col sep=comma, x expr=\coordindex, y expr={\thisrow{mean}+1*\thisrow{std}}]{\qamsf};
\addplot [
    color=color3,
    opacity=0.25,
    forget plot,
    ] fill between [
    of=A and B
];
\addplot[
        color=color3,
        semithick,
        mark=triangle
] table [x expr=\coordindex,y=mean, col sep=comma]{\qamsf};

\addplot[name path = A, draw=none, forget plot] 
    table [col sep=comma, x expr=\coordindex, y expr={\thisrow{mean}-1*\thisrow{std}}]{\ofdm};
\addplot[name path = B, draw=none, forget plot] 
    table [col sep=comma, x expr=\coordindex, y expr={\thisrow{mean}+1*\thisrow{std}}]{\ofdm};
\addplot [
    color=color4,
    opacity=0.25,
    forget plot,
    ] fill between [
    of=A and B
];
\addplot[
        color=color4,
        semithick,
        mark=diamond,
] table [x expr=\coordindex,y=mean, col sep=comma]{\ofdm};

\end{axis} 
\end{tikzpicture}%
			\end{minipage}
		}%
		\def\ylabel{Standardized feature value}%
		\def\path{tikz/dependency/snr/c42/}%
		\def\feature{$\tilde{C}_{42}$}%
		\tikzsetnextfilename{dependency_snr_c42}%
		\subfloat[][Dependency of $\tilde{C}_{42}$ on SNR]{%
			\begin{minipage}[t]{0.5\columnwidth}
				\centering

\global\edef\bpsk{\path BPSK.csv}
\global\edef\qpsk{\path QPSK.csv}
\global\edef\qamst{\path 16QAM.csv}
\global\edef\qamsf{\path 64QAM.csv}
\global\edef\ofdm{\path OFDM.csv}

\ifthenelse{\boolean{snr}}
{
    \global\edef\labels{{20dB, 15dB, 10dB, 5dB, 0dB}}
    \global\edef\xlabel{SNR class}
    \global\edef\xmax{4}
}{
    \global\edef\labels{{20dB, 15dB, 10dB, 5dB}}
    \global\edef\xlabel{SIR class}
    \global\edef\xmax{3}
}

\begin{tikzpicture} 

\begin{axis}[
    set layers,
    every axis plot/.append style={on layer=pre main},
    width=\figurewidth,
    height=\figureheight,
    title style={yshift=-7},
    xlabel=\xlabel,
    ylabel=\ylabel,
    xtick = {0,1,2,3,4},
    xticklabels/.expanded = {\labels},
    xticklabel style={font=\scriptsize},
    yticklabel style={font=\scriptsize},
    ylabel style={yshift=-3pt},
    ytick={-5,-4,-3,-2,-1,0,1,2,3,4,5},
    xmin=0, xmax=\xmax,
    clip mode=individual,
    clip marker paths=true,
    font=\footnotesize,
]

\addplot[name path = A, draw=none, forget plot] 
    table [col sep=comma, x expr=\coordindex, y expr={\thisrow{mean}-1*\thisrow{std}}]{\bpsk};
\addplot[name path = B, draw=none, forget plot] 
    table [col sep=comma, x expr=\coordindex, y expr={\thisrow{mean}+1*\thisrow{std}}]{\bpsk};
\addplot [
    color=color0,
    opacity=0.25,
    forget plot,
    ] fill between [
    of=A and B
];
\addplot[
        color=color0,
        semithick,
        mark=x
] table [x expr=\coordindex,y=mean, col sep=comma]{\bpsk};

\addplot[name path = A, draw=none, forget plot] 
    table [col sep=comma, x expr=\coordindex, y expr={\thisrow{mean}-1*\thisrow{std}}]{\qpsk};
\addplot[name path = B, draw=none, forget plot] 
    table [col sep=comma, x expr=\coordindex, y expr={\thisrow{mean}+1*\thisrow{std}}]{\qpsk};
\addplot [
    color=color1,
    opacity=0.25,
    forget plot,
    ] fill between [
    of=A and B
];
\addplot[
        color=color1,
        semithick,
        mark=square
] table [x expr=\coordindex,y=mean, col sep=comma]{\qpsk};

\addplot[name path = A, draw=none, forget plot] 
    table [col sep=comma, x expr=\coordindex, y expr={\thisrow{mean}-1*\thisrow{std}}]{\qamst};
\addplot[name path = B, draw=none, forget plot] 
    table [col sep=comma, x expr=\coordindex, y expr={\thisrow{mean}+1*\thisrow{std}}]{\qamst};
\addplot [
    color=color2,
    opacity=0.25,
    forget plot,
    ] fill between [
    of=A and B
];
\addplot[
        color=color2,
        semithick,
        mark=o
] table [x expr=\coordindex,y=mean, col sep=comma]{\qamst};

\addplot[name path = A, draw=none, forget plot] 
    table [col sep=comma, x expr=\coordindex, y expr={\thisrow{mean}-1*\thisrow{std}}]{\qamsf};
\addplot[name path = B, draw=none, forget plot] 
    table [col sep=comma, x expr=\coordindex, y expr={\thisrow{mean}+1*\thisrow{std}}]{\qamsf};
\addplot [
    color=color3,
    opacity=0.25,
    forget plot,
    ] fill between [
    of=A and B
];
\addplot[
        color=color3,
        semithick,
        mark=triangle
] table [x expr=\coordindex,y=mean, col sep=comma]{\qamsf};

\addplot[name path = A, draw=none, forget plot] 
    table [col sep=comma, x expr=\coordindex, y expr={\thisrow{mean}-1*\thisrow{std}}]{\ofdm};
\addplot[name path = B, draw=none, forget plot] 
    table [col sep=comma, x expr=\coordindex, y expr={\thisrow{mean}+1*\thisrow{std}}]{\ofdm};
\addplot [
    color=color4,
    opacity=0.25,
    forget plot,
    ] fill between [
    of=A and B
];
\addplot[
        color=color4,
        semithick,
        mark=diamond,
] table [x expr=\coordindex,y=mean, col sep=comma]{\ofdm};

\end{axis} 
\end{tikzpicture}%
			\end{minipage}
		}				
	\caption{SNR dependency of two example features, for the \textit{baseline} scenario (shaded regions indicate std. dev.). Unlike the SNR-corrected cumulant~$\tilde{C}_{42}$, $\sigma_{a}$ increases with decreasing SNR for SC-modulated signals as their amplitude distributions approach that of pure noise.}
	\label{fig:dependency}
\end{figure}
\tikzexternaldisable

\subsection{Effect of Transceiver Hardware}

\begin{table}[b!]
	\centering
			\caption{Classification accuracy ($\pm$ std. dev.) for combinations of transceiver hardware used in training and testing datasets}		
	\label{tab:hardware}
	\begingroup
		\aboverulesep=0ex
		\belowrulesep=0ex
		\footnotesize
		\setlength{\tabcolsep}{2pt} 
		\begin{tabularx}{\columnwidth}{p{1.2cm}|>{\centering\arraybackslash}X>{\centering\arraybackslash}X>{\centering\arraybackslash}X>{\centering\arraybackslash}X>{\centering\arraybackslash}X>{\centering\arraybackslash}X}
		\toprule
		\diagbox[innerwidth=1.2cm, height=0.8cm, font=\scriptsize]{\parbox[t][0.5cm]{0.8cm}{\linespread{0.9}\bf Test\\\bf/Train}}{\raisebox{-0.2cm}{\bf SNR}} &  \SI{0}{dB} & \SI{5}{dB} & \SI{10}{dB} & \SI{15}{dB} & \SI{20}{dB} & \parbox{1cm}{Average accuracy} \\
		\midrule
		\rule{0pt}{0.3cm}\multirow{ 2}{*}[2pt]{Lab/Lab} & 90.4\% & 98.8\% & 100.0\% & 100.0\% & 100.0\% & 97.8\% \\[-0.8ex]
		& \tiny$(\pm0.5\%)$ & \tiny$(\pm0.2\%)$ & \tiny$(\pm0.0\%)$ & \tiny$(\pm0.0\%)$ & \tiny$(\pm0.0\%)$ & \tiny$(\pm0.1\%)$ \\[0.4ex]
		\multirow{ 2}{*}[2pt]{SDR/SDR} & 90.4\% & 98.4\% & 100.0\% & 100.0\% & 100.0\% & 97.8\% \\[-0.8ex]
		& \tiny$(\pm0.5\%)$ & \tiny$(\pm0.3\%)$ & \tiny$(\pm0.0\%)$ & \tiny$(\pm0.0\%)$ & \tiny$(\pm0.0\%)$ & \tiny$(\pm0.1\%)$ \\[0.4ex]
		\multirow{ 2}{*}[2pt]{SDR/Lab} & 86.2\% & 96.0\% & 93.2\% & 93.3\% & 97.3\% & 93.2\% \\[-0.8ex]
		& \tiny$(\pm0.6\%)$ & \tiny$(\pm0.6\%)$ & \tiny$(\pm1.5\%)$ & \tiny$(\pm1.2\%)$ & \tiny$(\pm0.7\%)$ & \tiny$(\pm0.8\%)$ \\[0.4ex]
		\multirow{ 2}{*}[2pt]{Lab/SDR} & 90.3\% & 97.1\% & 99.7\% & 99.1\% & 81.6\% & 93.6\% \\[-0.8ex]
		& \tiny$(\pm0.6\%)$ & \tiny$(\pm0.8\%)$ & \tiny$(\pm0.3\%)$ & \tiny$(\pm0.6\%)$ & \tiny$(\pm10.1\%)$ & \tiny$(\pm2.1\%)$ \\
		\bottomrule
		\end{tabularx}
	\endgroup
\end{table}

Table~\ref{tab:hardware} shows the results from four experiments of running the classifier on combinations of training and testing datasets, generated using different transceiver hardware as per Section~\ref{scen:sdr}. As evident from the first two rows, the accuracy of the linear classifier does not degrade significantly if we use SDRs instead of laboratory-grade hardware as transceivers for both training and testing. Even at low SNR levels, i.e. $\SI{0}{dB}$ and $\SI{5}{dB}$, the accuracy is nearly the same for both device classes. This shows that, once the SNR is accounted for, the classifier is fairly agnostic towards the signal quality, as long as the training and testing datasets were recorded using the same device type. However, the bottom two rows of Table~\ref{tab:hardware} show that the classifier performance degrades when we use datasets recorded from dissimilar hardware during training and testing. The average classification accuracy decreases by about $4.5\%$ and, interestingly, estimates especially in the high-SNR domain become inaccurate. One reason for this may be that the higher noise figure of the SDRs necessitates a higher signal energy for a given SNR, and that some of the features, though normalized, do not generalize well under these circumstances. 

Therefore, Table~\ref{tab:hardware} shows that it is not inherently challenging to run DMC on lower-integrity RF signals at the range of SNRs typical for modern wireless systems. However, the transceiver hardware does have an influence on the distribution of the features once the SNR is accounted for, which leads to confusions during classification if the training dataset does not capture this hardware variety. Our results thus suggest that it is not straightforward to generalize results from one hardware test to a wider set of radios, nor to train the classifier using one radio type and seamlessly run DMC for other devices.

\subsection{Effect of Co-Channel Interference}

\begin{table}[tb!]
	\centering
	\caption{Classification accuracy ($\pm$ std. dev.) under co-channel interference during training and testing, at SNR=\SI{10}{dB}}
	\label{tab:intf_train}
	\begingroup
		\aboverulesep=0ex
		\belowrulesep=0ex
		\footnotesize
		\setlength{\tabcolsep}{2pt} 
		\begin{tabularx}{\columnwidth}{p{2cm}|>{\centering\arraybackslash}X>{\centering\arraybackslash}X>{\centering\arraybackslash}X>{\centering\arraybackslash}X>{\centering\arraybackslash}X>{\centering\arraybackslash}X}
		\toprule
		\diagbox[innerwidth=2cm, height=0.8cm, font=\scriptsize]{\parbox[t][0.5cm]{2cm}{\linespread{0.9}\bf Inter-\\\bf fering signal}}{\raisebox{-0.225cm}{\bf SIR}} & \SI{5}{dB} & \SI{10}{dB} & \SI{15}{dB} & \SI{20}{dB} & \parbox{1cm}{Average accuracy} \\
		\midrule
		\rule{0pt}{0.3cm}\multirow{ 2}{*}[2pt]{SC BPSK} & 97.4\% & 98.7\% & 98.2\% & 98.0\% & 97.8\% \\[-0.8ex]
		& \tiny$(\pm0.3\%)$ & \tiny$(\pm0.1\%)$ & \tiny$(\pm0.2\%)$ & \tiny$(\pm0.2\%)$ & \tiny$(\pm0.1\%)$ \\[0.4ex]
		\multirow{ 2}{*}[2pt]{SC 16-QAM} & 90.6\% & 92.7\% & 99.1\% & 98.3\% & 95.9\%\\[-0.8ex]
		& \tiny$(\pm0.4\%)$ & \tiny$(\pm0.4\%)$ & \tiny$(\pm0.2\%)$ & \tiny$(\pm0.2\%)$ & \tiny$(\pm0.1\%)$ \\[0.4ex]
		\multirow{ 2}{*}[2pt]{OFDM 64-QAM} & 93.0\% & 96.5\% & 98.2\% & 99.8\% & 97.4\%\\[-0.8ex]
		& \tiny$(\pm0.2\%)$ & \tiny$(\pm0.2\%)$ & \tiny$(\pm0.2\%)$ & \tiny$(\pm0.1\%)$ & \tiny$(\pm0.1\%)$ \\
		\bottomrule
		\end{tabularx}
	\endgroup
\end{table}

\begin{table}[tb!]
	\centering
	\caption{Classification accuracy ($\pm$ std. dev.) under co-channel interference at SNR=\SI{10}{dB}, for training without interference}
	\label{tab:intf_test}
	\begingroup
		\aboverulesep=0ex
		\belowrulesep=0ex
		\footnotesize
		\setlength{\tabcolsep}{2pt} 
		\begin{tabularx}{\columnwidth}{p{2cm}|>{\centering\arraybackslash}X>{\centering\arraybackslash}X>{\centering\arraybackslash}X>{\centering\arraybackslash}X>{\centering\arraybackslash}X>{\centering\arraybackslash}X}
		\toprule
		\diagbox[innerwidth=2cm, height=0.8cm, font=\scriptsize]{\parbox[t][0.5cm]{2cm}{\linespread{0.9}\bf Inter-\\\bf fering signal}}{\raisebox{-0.225cm}{\bf SIR}} & \SI{5}{dB} & \SI{10}{dB} & \SI{15}{dB} & \SI{20}{dB} & \parbox{1cm}{Average accuracy} \\
		\midrule
		\rule{0pt}{0.3cm}\multirow{ 2}{*}[2pt]{SC BPSK} & 39.8\% & 70.1\% & 80.7\% & 94.1\% & 71.2\% \\[-0.8ex]
		& \tiny$(\pm0.5\%)$ & \tiny$(\pm2.5\%)$ & \tiny$(\pm0.3\%)$ & \tiny$(\pm1.0\%)$ & \tiny$(\pm0.7\%)$ \\[0.4ex]
		\multirow{ 2}{*}[2pt]{SC 16-QAM} & 40.5\% & 59.4\% & 80.6\% & 98.4\% & 69.7\%\\[-0.8ex]
		& \tiny$(\pm0.5\%)$ & \tiny$(\pm2.5\%)$ & \tiny$(\pm0.5\%)$ & \tiny$(\pm0.5\%)$ & \tiny$(\pm0.7\%)$ \\[0.4ex]
		\multirow{ 2}{*}[2pt]{OFDM 64-QAM} & 40.1\% & 59.3\% & 80.6\% & 94.9\% & 68.7\%\\[-0.8ex]
		& \tiny$(\pm0.4\%)$ & \tiny$(\pm2.2\%)$ & \tiny$(\pm0.5\%)$ & \tiny$(\pm0.6\%)$ & \tiny$(\pm0.6\%)$ \\
		\bottomrule
		\end{tabularx}
	\endgroup
\end{table}


In this section we present the results for the modulated co-channel interference scenario in Section~\ref{scen:cci}. Table~\ref{tab:intf_train} shows the classification performance at $\text{SNR}=\SI{10}{dB}$ with interference of selected modulations over a range of SIR values being present both during testing and training. The results show that DMC accuracy generally degrades by a few percent. 
We have confirmed that all misclassifications in Table~\ref{tab:intf_train} stem from the difficulty of differentiating between 16-QAM and 64-QAM. This is consistent with SC 16-QAM interference confusing the classifier the most out of the three tested modulations, degrading the accuracy by approximately $10\%$ at an SIR of \SI{5}{dB} in comparison to the no-interference baseline. Nonetheless, the average accuracy is above 95\%, even in the presence of interference. 

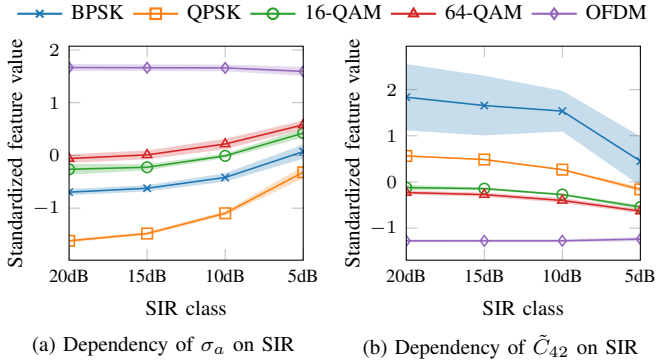
\begin{figure}[tb!]
	\centering
		\setboolean{snr}{false}%
		\setlength\figureheight{0.5\columnwidth}%
		\setlength\figurewidth{0.53\columnwidth}%
		\subfloat{

\begin{tikzpicture} 
    \begin{axis}[%
        hide axis,
        xmin=0, xmax=1,
        ymin=0, ymax=1,
        font=\footnotesize,
        legend columns=5,
        legend style={draw=none, column sep=1pt},
    ]
    \addlegendimage{color0, thick, mark=x}
    \addlegendimage{color1, thick, mark=square}
    \addlegendimage{color2, thick, mark=o}
    \addlegendimage{color3, thick, mark=triangle}
    \addlegendimage{color4, thick, mark=diamond}
    \addlegendentry{BPSK}
    \addlegendentry{QPSK}
    \addlegendentry{16-QAM}
    \addlegendentry{64-QAM}
    \addlegendentry{OFDM}
    \end{axis}
\end{tikzpicture}%
		}%
		\vspace{-0.4cm}%
		\setcounter{subfigure}{0}%
		\def\ylabel{Standardized feature value}
		\def\path{tikz/dependency/sir/sigma_a/}
		\tikzsetnextfilename{dependency_sir_sigmaa}%
		\subfloat[][Dependency of $\sigma_{a}$ on SIR]{%

			\begin{minipage}[t]{0.48\columnwidth}
				\centering

\global\edef\bpsk{\path BPSK.csv}
\global\edef\qpsk{\path QPSK.csv}
\global\edef\qamst{\path 16QAM.csv}
\global\edef\qamsf{\path 64QAM.csv}
\global\edef\ofdm{\path OFDM.csv}

\ifthenelse{\boolean{snr}}
{
    \global\edef\labels{{20dB, 15dB, 10dB, 5dB, 0dB}}
    \global\edef\xlabel{SNR class}
    \global\edef\xmax{4}
}{
    \global\edef\labels{{20dB, 15dB, 10dB, 5dB}}
    \global\edef\xlabel{SIR class}
    \global\edef\xmax{3}
}

\begin{tikzpicture} 

\begin{axis}[
    set layers,
    every axis plot/.append style={on layer=pre main},
    width=\figurewidth,
    height=\figureheight,
    title style={yshift=-7},
    xlabel=\xlabel,
    ylabel=\ylabel,
    xtick = {0,1,2,3,4},
    xticklabels/.expanded = {\labels},
    xticklabel style={font=\scriptsize},
    yticklabel style={font=\scriptsize},
    ylabel style={yshift=-3pt},
    ytick={-5,-4,-3,-2,-1,0,1,2,3,4,5},
    xmin=0, xmax=\xmax,
    clip mode=individual,
    clip marker paths=true,
    font=\footnotesize,
]

\addplot[name path = A, draw=none, forget plot] 
    table [col sep=comma, x expr=\coordindex, y expr={\thisrow{mean}-1*\thisrow{std}}]{\bpsk};
\addplot[name path = B, draw=none, forget plot] 
    table [col sep=comma, x expr=\coordindex, y expr={\thisrow{mean}+1*\thisrow{std}}]{\bpsk};
\addplot [
    color=color0,
    opacity=0.25,
    forget plot,
    ] fill between [
    of=A and B
];
\addplot[
        color=color0,
        semithick,
        mark=x
] table [x expr=\coordindex,y=mean, col sep=comma]{\bpsk};

\addplot[name path = A, draw=none, forget plot] 
    table [col sep=comma, x expr=\coordindex, y expr={\thisrow{mean}-1*\thisrow{std}}]{\qpsk};
\addplot[name path = B, draw=none, forget plot] 
    table [col sep=comma, x expr=\coordindex, y expr={\thisrow{mean}+1*\thisrow{std}}]{\qpsk};
\addplot [
    color=color1,
    opacity=0.25,
    forget plot,
    ] fill between [
    of=A and B
];
\addplot[
        color=color1,
        semithick,
        mark=square
] table [x expr=\coordindex,y=mean, col sep=comma]{\qpsk};

\addplot[name path = A, draw=none, forget plot] 
    table [col sep=comma, x expr=\coordindex, y expr={\thisrow{mean}-1*\thisrow{std}}]{\qamst};
\addplot[name path = B, draw=none, forget plot] 
    table [col sep=comma, x expr=\coordindex, y expr={\thisrow{mean}+1*\thisrow{std}}]{\qamst};
\addplot [
    color=color2,
    opacity=0.25,
    forget plot,
    ] fill between [
    of=A and B
];
\addplot[
        color=color2,
        semithick,
        mark=o
] table [x expr=\coordindex,y=mean, col sep=comma]{\qamst};

\addplot[name path = A, draw=none, forget plot] 
    table [col sep=comma, x expr=\coordindex, y expr={\thisrow{mean}-1*\thisrow{std}}]{\qamsf};
\addplot[name path = B, draw=none, forget plot] 
    table [col sep=comma, x expr=\coordindex, y expr={\thisrow{mean}+1*\thisrow{std}}]{\qamsf};
\addplot [
    color=color3,
    opacity=0.25,
    forget plot,
    ] fill between [
    of=A and B
];
\addplot[
        color=color3,
        semithick,
        mark=triangle
] table [x expr=\coordindex,y=mean, col sep=comma]{\qamsf};

\addplot[name path = A, draw=none, forget plot] 
    table [col sep=comma, x expr=\coordindex, y expr={\thisrow{mean}-1*\thisrow{std}}]{\ofdm};
\addplot[name path = B, draw=none, forget plot] 
    table [col sep=comma, x expr=\coordindex, y expr={\thisrow{mean}+1*\thisrow{std}}]{\ofdm};
\addplot [
    color=color4,
    opacity=0.25,
    forget plot,
    ] fill between [
    of=A and B
];
\addplot[
        color=color4,
        semithick,
        mark=diamond,
] table [x expr=\coordindex,y=mean, col sep=comma]{\ofdm};

\end{axis} 
\end{tikzpicture}%
			\end{minipage}
		}\hfill%
		\def\ylabel{Standardized feature value}%
		\def\path{tikz/dependency/sir/c42/}
		\def\feature{$\tilde{C}_{42}$}%
		\tikzsetnextfilename{dependency_sir_c42}%
		\subfloat[][Dependency of $\tilde{C}_{42}$ on SIR]{%
			\begin{minipage}[t]{0.48\columnwidth}
				\centering

\global\edef\bpsk{\path BPSK.csv}
\global\edef\qpsk{\path QPSK.csv}
\global\edef\qamst{\path 16QAM.csv}
\global\edef\qamsf{\path 64QAM.csv}
\global\edef\ofdm{\path OFDM.csv}

\ifthenelse{\boolean{snr}}
{
    \global\edef\labels{{20dB, 15dB, 10dB, 5dB, 0dB}}
    \global\edef\xlabel{SNR class}
    \global\edef\xmax{4}
}{
    \global\edef\labels{{20dB, 15dB, 10dB, 5dB}}
    \global\edef\xlabel{SIR class}
    \global\edef\xmax{3}
}

\begin{tikzpicture} 

\begin{axis}[
    set layers,
    every axis plot/.append style={on layer=pre main},
    width=\figurewidth,
    height=\figureheight,
    title style={yshift=-7},
    xlabel=\xlabel,
    ylabel=\ylabel,
    xtick = {0,1,2,3,4},
    xticklabels/.expanded = {\labels},
    xticklabel style={font=\scriptsize},
    yticklabel style={font=\scriptsize},
    ylabel style={yshift=-3pt},
    ytick={-5,-4,-3,-2,-1,0,1,2,3,4,5},
    xmin=0, xmax=\xmax,
    clip mode=individual,
    clip marker paths=true,
    font=\footnotesize,
]

\addplot[name path = A, draw=none, forget plot] 
    table [col sep=comma, x expr=\coordindex, y expr={\thisrow{mean}-1*\thisrow{std}}]{\bpsk};
\addplot[name path = B, draw=none, forget plot] 
    table [col sep=comma, x expr=\coordindex, y expr={\thisrow{mean}+1*\thisrow{std}}]{\bpsk};
\addplot [
    color=color0,
    opacity=0.25,
    forget plot,
    ] fill between [
    of=A and B
];
\addplot[
        color=color0,
        semithick,
        mark=x
] table [x expr=\coordindex,y=mean, col sep=comma]{\bpsk};

\addplot[name path = A, draw=none, forget plot] 
    table [col sep=comma, x expr=\coordindex, y expr={\thisrow{mean}-1*\thisrow{std}}]{\qpsk};
\addplot[name path = B, draw=none, forget plot] 
    table [col sep=comma, x expr=\coordindex, y expr={\thisrow{mean}+1*\thisrow{std}}]{\qpsk};
\addplot [
    color=color1,
    opacity=0.25,
    forget plot,
    ] fill between [
    of=A and B
];
\addplot[
        color=color1,
        semithick,
        mark=square
] table [x expr=\coordindex,y=mean, col sep=comma]{\qpsk};

\addplot[name path = A, draw=none, forget plot] 
    table [col sep=comma, x expr=\coordindex, y expr={\thisrow{mean}-1*\thisrow{std}}]{\qamst};
\addplot[name path = B, draw=none, forget plot] 
    table [col sep=comma, x expr=\coordindex, y expr={\thisrow{mean}+1*\thisrow{std}}]{\qamst};
\addplot [
    color=color2,
    opacity=0.25,
    forget plot,
    ] fill between [
    of=A and B
];
\addplot[
        color=color2,
        semithick,
        mark=o
] table [x expr=\coordindex,y=mean, col sep=comma]{\qamst};

\addplot[name path = A, draw=none, forget plot] 
    table [col sep=comma, x expr=\coordindex, y expr={\thisrow{mean}-1*\thisrow{std}}]{\qamsf};
\addplot[name path = B, draw=none, forget plot] 
    table [col sep=comma, x expr=\coordindex, y expr={\thisrow{mean}+1*\thisrow{std}}]{\qamsf};
\addplot [
    color=color3,
    opacity=0.25,
    forget plot,
    ] fill between [
    of=A and B
];
\addplot[
        color=color3,
        semithick,
        mark=triangle
] table [x expr=\coordindex,y=mean, col sep=comma]{\qamsf};

\addplot[name path = A, draw=none, forget plot] 
    table [col sep=comma, x expr=\coordindex, y expr={\thisrow{mean}-1*\thisrow{std}}]{\ofdm};
\addplot[name path = B, draw=none, forget plot] 
    table [col sep=comma, x expr=\coordindex, y expr={\thisrow{mean}+1*\thisrow{std}}]{\ofdm};
\addplot [
    color=color4,
    opacity=0.25,
    forget plot,
    ] fill between [
    of=A and B
];
\addplot[
        color=color4,
        semithick,
        mark=diamond,
] table [x expr=\coordindex,y=mean, col sep=comma]{\ofdm};

\end{axis} 
\end{tikzpicture}%
			\end{minipage}
		}				
	\caption{SIR dependency of two example features, for the \textit{interference} scenario at SNR=\SI{10}{dB} (shaded regions indicate std. dev.). Since the interference power is not estimated by the blind receiver, both features are SIR-dependent.}
	\label{fig:dependency_sir}
\end{figure}

Table~\ref{tab:intf_test} also shows the classification accuracy for the \emph{interference} scenario, but when the classifier has been trained solely on interference-free data of the same SNR level. Comparing the results in Table~\ref{tab:intf_train} and Table~\ref{tab:intf_test} demonstrates that training only on the interference-free dataset has a highly detrimental effect on the classification accuracy. 
Analogously to Fig.~\ref{fig:dependency}, Fig.~\ref{fig:dependency_sir} shows the effect of SIR on two features at the fixed SNR of \SI{10}{dB}, illustrating that both features are SIR-dependent. We note that in this scenario, the SNR estimate $\hat{SNR}$ used to correct the cumulant feature values becomes biased, because the blind receiver cannot differentiate between the signal and in-band interference in the power domain. Further analysis of the other features has confirmed that they are also SIR-dependent. It thus follows that for any classifier without a model for feature-dependency on SIR, it is difficult to accurately classify interfered signals if the interference was not also present at training time.

\tikzexternalenable
\tikzexternaldisable

\section{Conclusions}\label{sec:conclusion}
In this paper, we examined the performance of DMC for RF signals in real-world wireless networks. We used a linear SVM classifier with twelve widely-adopted signal features, and tested the classification accuracy on an extensive RF signal dataset of \num[group-separator={,}]{250000} over-the-air transmissions. Our experiments encompassed different noise levels, heterogeneous transceiver hardware, and modulated co-channel interference. Our results exhibit a high classification accuracy even when tested under the challenging real-world conditions of modulated co-channel interference and low-grade hardware, if the training dataset is representative of these impairments. However, the DMC performance deteriorates significantly if the training set does not properly capture the variety of the real radio environment, in terms of interference and hardware types. We emphasize that the performance degradation is not an issue of overfitting to the training data, but rather a problem of bias inherent to the features. Our results have two important engineering implications. First, they suggest that it may not be feasible to easily exchange learned classifier models among dissimilar radio environments and devices in practice, and that comprehensive datasets are required for robust training. Second, our work suggests that the research community should focus strongly on designing features that generalize well to new radio scenarios as the key missing link for the real-world deployment of DMC, rather than applying new complex machine learning algorithms.  We are making our dataset publicly available\cite{database}, in a step towards a unified RF database for evaluating future DMC algorithms in a standardized and realistic framework.

\tikzexternaldisable


\bibliography{IEEEfull,bibliography}
\bibliographystyle{IEEE}   

\section*{Appendix - Selected Feature Definitions}

\begingroup
\footnotesize
\begin{equation}\label{eq:snr}
\hat{SNR} = \left(\textstyle\int_B P(f)df-\textstyle\int_N P(f)df\right)/\textstyle\int_N P(f)df,
\end{equation}
with the periodogram $P(f)$ of the signal integrated over the signal band $B$ and an equally wide band $N$ ($B\cap N = \emptyset$) for noise power estimation. The SNR estimate is biased once interference is introduced.






\begin{equation}\label{eq:nw}
N_c=\lvert \{n~\vert~A_{N}[n]>A_t\} \rvert,
\end{equation}
with $A_t=1$ and $A_{N}[n] = A[n]/\mu_A$.

\begin{equation}\label{eq:c42}
    \tilde{C}_{42} = \tfrac{C_{42}}{\left(P_s-\sigma_N^2\right)^2}=C_{42}\left(\tfrac{1+\hat{SNR}}{P_s\hat{SNR}}\right)^2,
\end{equation} 
\begin{equation} \label{eq:c63}
    \tilde{C}_{63} = \tfrac{C_{63}}{\left(P_s-\sigma_N^2\right)^3}=C_{63}\left(\tfrac{1+\hat{SNR}}{P_s\hat{SNR}}\right)^3,
\end{equation} 
with $C_{42}=cum(s,s,s^*,s^*)$~\cite{swami2000}, $C_{63}=cum(s,s,s,s^*,s^*,s^*)$~\cite{orlic2012}, the power of the received signal $P_s$ and the noise variance $\sigma_N^2$.



\endgroup



\ifpreprint
	\newpage
	\onecolumn
	$\copyright$ 2018 IEEE. Personal use of this material is permitted. Permission from IEEE must be obtained for all other uses, in any current or future media, including reprinting/republishing this material for advertising or promotional purposes, creating new collective works, for resale or redistribution to servers or lists, or reuse of any copyrighted component of this work in other works.
	
	\bigskip
	
	This work will be published in:

	C.~{de Vrieze}, L.~Simi\'c, and P.~M\"ah\"onen, ``The importance of being earnest: Performance of modulation classification for real {RF} signals,'' in \emph{IEEE International Symposium on Dynamic Spectrum Access Networks (IEEE DySPAN 2018)}, Seoul, 2018.

\else
\fi 

\end{document}